\documentclass[aps,pra,twocolumn,superscriptaddress,floatfix,footinbib,notitlepage,groupaddress,showpacs]{revtex4-1} 
\usepackage{times,color,amsfonts,amssymb,amsbsy,amsthm,amsmath,graphics,graphicx,hyperref,bm,bbm,dsfont,hyperref,mathrsfs,color,xcolor,cancel,float}
\newcommand{\ignore}[1]{}

\hypersetup{colorlinks,linkcolor={blue},citecolor={blue},urlcolor={blue}}
\newcommand{\ket}[1]{\left|{#1}\right\rangle}
\newcommand{\bra}[1]{\left\langle{#1}\right|}

\DeclareFontFamily{OT1}{pzc}{}
\DeclareFontShape{OT1}{pzc}{m}{it}%
              {<-> s * [1.25] pzcmi7t}{}
\DeclareMathAlphabet{\mathpzc}{OT1}{pzc}%
                                 {m}{it}

\let\oldsqrt\sqrt
\def\sqrt{\mathpalette\DHLhksqrt}
\def\DHLhksqrt#1#2{%
\setbox0=\hbox{$#1\oldsqrt{#2\,}$}\dimen0=\ht0
\advance\dimen0-0.2\ht0
\setbox2=\hbox{\vrule height\ht0 depth -\dimen0}%
{\box0\lower0.4pt\box2}}

\begin{document}

\title{Fluctuation relation for heat exchange in Markovian open quantum systems}

\author{M. Ramezani}
\affiliation{Department of Physics, Sharif University of Technology, Tehran 14588, Iran}
\affiliation{School of Physics, Institute for Research in Fundamental Sciences (IPM), Tehran 19395, Iran}

\author{M. Golshani}
\affiliation{Department of Physics, Sharif University of Technology, Tehran 14588, Iran}
\affiliation{School of Physics, Institute for Research in Fundamental Sciences (IPM), Tehran 19395, Iran}

\author{A. T. Rezakhani}
\affiliation{Department of Physics, Sharif University of Technology, Tehran 14588, Iran}

\date{\today}

\begin{abstract}
A fluctuation relation for the heat exchange of an open quantum system under a thermalizing Markovian dynamics is derived. We show that the probability of that the system absorbs an amount of heat from its bath, at a given time interval, divided by the probability of the reverse process (releasing the same amount of heat to the bath) is given by an exponential factor which depends on the amount of heat and the difference between the temperatures of the system and the bath. We also argue that the probability of the violation of the second law of thermodynamics (here in the form of net heat transfer from a cold system to its hot bath) drops exponentially with both the amount of heat and the temperature differences.
\end{abstract}

\maketitle

\section{Introduction}
\label{sec:intr}

Irreversibility of dynamics is a ubiquitous feature of macroscopic systems, which appears despite microscopic reversibility in classical and quantum systems \cite{book:thermodynamics,Crooks:irrev,book:Breuer-Petruccione,book:Rivas-Huelga}. The relation between these macroscopic and microscopic features can be captured by ``fluctuation relations" (FRs) \cite{1997-Jarzynski,1999-Crooks,2000-Crooks,2002-Evans,2004-Jarzynski,Hanggi-RMP}. For example, for a Markovian process, where the initial and final states of the system are thermal (with inverse temperature $\beta=1/T$, taking the Boltzmann constant $k_{B}\equiv 1$), the probabilities of doing work $\mathbbmss{W}$ on the system in the \textit{forward} path is related to the corresponding probability of doing work $-\mathbbmss{W}$ in the \textit{reverse} path, through \cite{1999-Crooks} 
\begin{equation}
\frac{P_{\textsc{f}}(+\mathbbmss{W})}{P_{\textsc{r}}(-\mathbbmss{W})}=e^{\beta (\mathbbmss{W}-\Delta \mathbbmss{F})},
\label{eq:crooks}
\end{equation}  
where $\Delta \mathbbmss{F}$ is the difference in the free energy of the system. A similar relation has been shown to govern heat exchange $\mathbbmss{Q}$ in the forward and reverse paths between two weakly-coupled systems $S$ and $B$ (from $B$ to $S$), with the difference $\Delta \beta= \beta_{S}-\beta_{B}$ in their inverse temperatures \cite{2004-Jarzynski,Abe,Akagawa,Goold,Landi},
\begin{equation}
\frac{P_{\textsc{f}}(+\mathbbmss{Q})}{P_{\textsc{r}}(-\mathbbmss{Q})}=e^{\mathbbmss{Q} \Delta \beta},
\end{equation}  

The distribution function of any quantity in the forward and reverse paths depends on two factors. One is the probability of initial state of the system, which is determined by preparation; the other is the probability of the path, which is determined by dynamics. For a closed system, due to the time reversibility of dynamics, the path probabilities are equal in the forward and reverse cases such that they cancel out each other when we calculate the ratio of the distributions \cite{2004-Jarzynski}. As a result, dynamics seems to play no explicit role in deriving FRs.

Dynamics of an open system is not necessarily time reversible, so the path probabilities in the forward and reverse paths are not the same. But in what follows we show that a similar FR (in some sense) is attainable. A principal question for open systems is what a \textit{reverse} dynamics physically mean.  Although there has been some progress towards defining reverse dynamics \cite{2008-Crooks,2015-Aurell,Lidar-etal}), an unambiguous definition has been elusive thus far. In order to avoid this issue, here we simply replace the notion of reverse \textit{dynamics} with reverse \textit{process}. If in the forward dynamics, the system releases heat $\mathbbmss{Q}$, the reverse process corresponds to absorbing heat $\mathbbmss{Q}$. This setting helps us express our finding fully in terms of the forward path. 
    
For a system $S$ prepared in a thermal state of temperature $T_{S}=1/\beta_{S}$, and then put in contact for time $\tau$ with a heat bath of temperature $T_{B}=1/\beta_{B}$, where the dynamics of the system is given by a thermalizing Lindblad equation (i.e., the dynamics drives the system to become thermal with the bath in a sufficiently long time), we can show that the following fluctuation relation holds:
\begin{equation}
\frac{P_{\textsc{f}}(+\mathbbmss{Q},\tau)}{P_{\textsc{f}}(-\mathbbmss{Q},\tau)}=e^{\mathbbmss{Q}\Delta \beta},
\label{Fluctuation_Relation}
\end{equation}
where $P_{\textsc{f}}(\pm \mathbbmss{Q},\tau)$ is the probability that the system \textit{absorbs} (\textit{releases}) heat $\mathbbmss{Q}$ from (to) the heat bath in the time interval $\tau$ in the forward path. We omit ``$\textsc{f}$" hereon in order to lighten the notation. In the following sections, after reviewing a fairly general model for thermalizing Markovian dynamics, we give the proof of our FR.

\section{Thermalizing dynamics of an open quantum system}
\label{sec:prelim}

\begin{figure*}[tp]
\includegraphics[scale=.33]{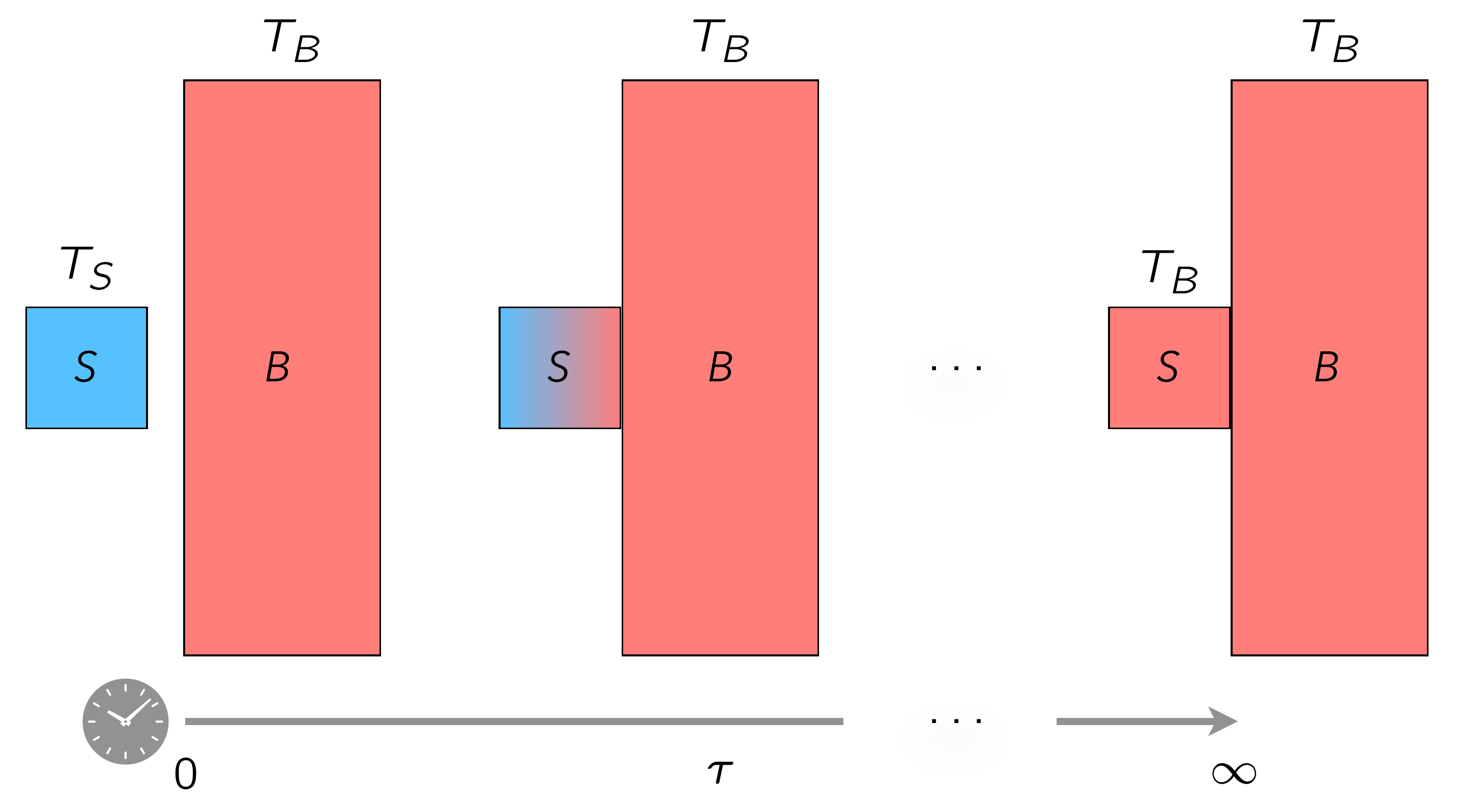} \hskip25mm \includegraphics[scale=.33]{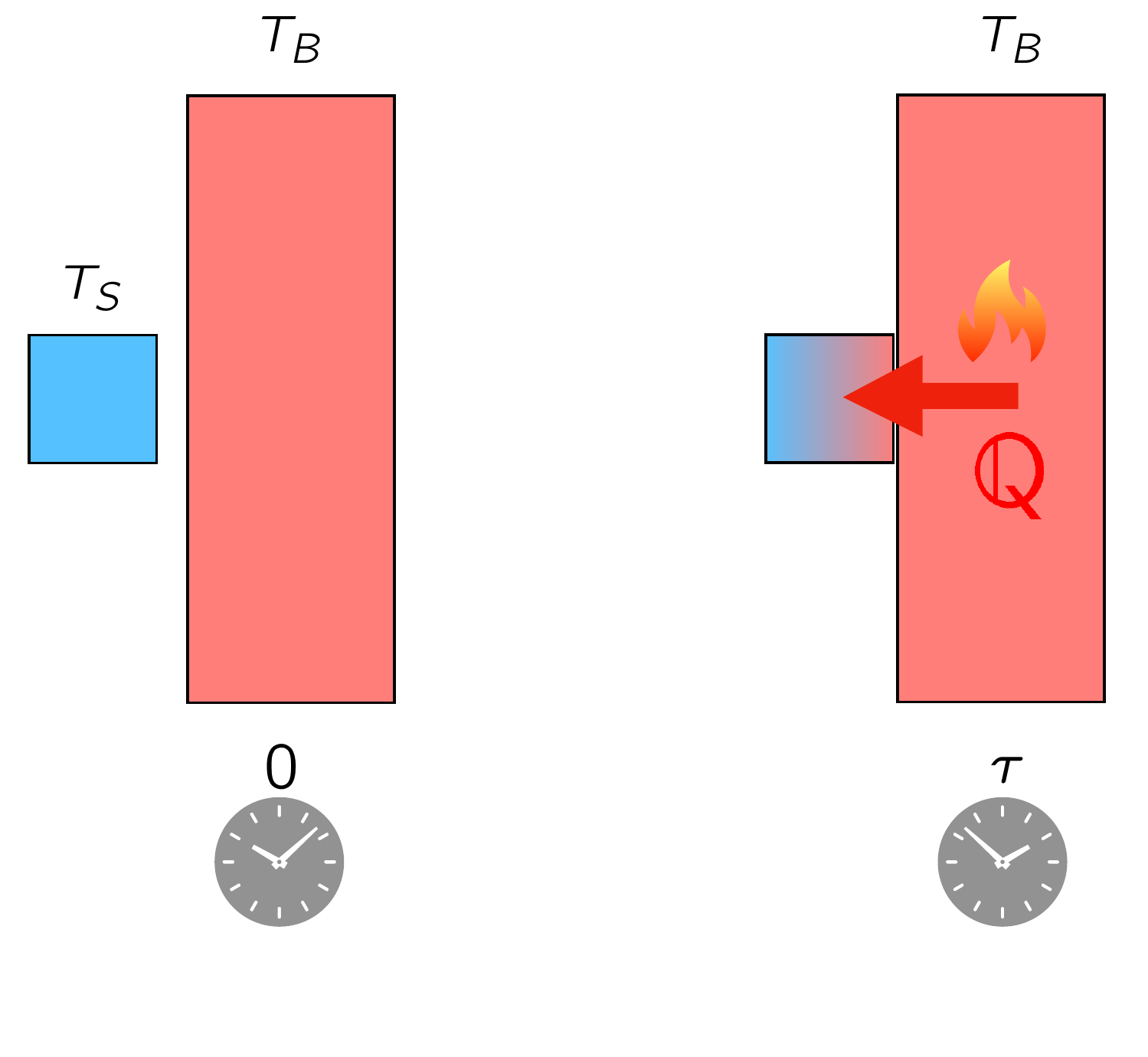} 
\caption{(Color online.) Left: Schematic of thermalization of system $S$ (with initial temperature $T_{S}$) in contact with a bath $B$ with temperature $T_{B}$. Right: Schematic of the process where at time $\tau$ the system $S$ absorbs heat $\mathbbmss{Q}$ from the bath $B$.}
\label{fig:scheme}
\end{figure*}

Consider a quantum system with a $D$-dimensional Hilbert space, and let $H_{0}$ be its free Hamiltonian and the set $\{|m\rangle\}$ indicate the eigenvectors corresponding to the discrete and nondegenerate eigenvalues $E^{(0)}_{m}$ ($H_{0}|m\rangle = E^{(0)}_{m} |m\rangle$), ordered increasingly as $E^{(0)}_{1}<E^{(0)}_{2}< \ldots <E^{(0)}_{D}$. Assume that the system is then put in contact with a heat bath (reservoir or environment) of inverse temperature $\beta_{B}$ and assume that the dynamics of the system is described (within the weak-coupling and Markovian approximation) by the Lindblad equation \cite{1976-Lindblad,review-Pascazio}
\begin{equation}
\label{Lindblad}
\frac{\mathrm{d}\varrho}{\mathrm{d}\tau}=-i[H,\varrho] + \sum_{a} \big( L_{a}\varrho L_{a}^{\dagger}-\frac{1}{2}\{L_{a}^{\dagger} L_{a},\varrho\} \big),
\end{equation}  
where $\varrho(\tau)$ is the density matrix of the system, $H$ representing an effective Hamiltonian (usually different from $H_{0}$), and $L_{a}$s are the quantum jump operators induced by the interaction with the bath, and we have set $\hbar\equiv 1$. 

Conditions under which a Lindbladian dynamics can yield a stationary state which is a thermal equilibrium state in a Gibbsian form have been studied in the literature extensively \cite{Spohn,book:Breuer-Petruccione,book:Rivas-Huelga,2016-Ostilli,Us}. Here we follow the formalism and context laid out recently in Ref. \cite{2016-Ostilli}. Here the conditions for the evolution (\ref{Lindblad})---with $L_{\alpha}$s chosen as $L_{mn}=l_{mn}|m\rangle\langle n|$ ($m\neq n$)---to have the stationary solution of the thermal form $\varrho^{(\mathrm{eq})}=e^{-\beta_{B} H_{0}}/\mathrm{Tr}[e^{-\beta_{B} H_{0}}]$ have been shown to be 
\begin{align}
\label{constraint_H}
H&=\sum_{m}E_{m}|m\rangle\langle m|,\\
\label{constraint_L}
|l_{mn}|^{2}&=C_{mn}e^{-\beta_{B}(E_{m}^{(0)}-E_{n}^{(0)})/2},
\end{align} 
where $C_{mn}=C_{nm}>0$ and $C_{mn}$s depend on the interaction of the system and the heat bath. From these relations it follows that the diagonal elements of $\varrho$ have an evolution decoupled from that of off-diagonal terms \cite{2016-Ostilli}. Defining the vector $|\mathbf{v}(\tau)\rangle=\sum_{m}\varrho_{mm}(\tau)|m\rangle$, we have the following evolution:
\begin{equation}
|\mathbf{v}(\tau)\rangle=e^{-\tau A} |\mathbf{v}(0)\rangle,
\end{equation}
where $A=\sum_{mn}A_{mn}|m\rangle\langle n|$ is defined as 
\begin{equation}
\label{Matrix_A}
A_{mn} = \begin{cases} 
\sum_{j\neq m}|l_{jm}|^2 ;      &m=n, \\
- |l_{mn}|^2  ;&  m\neq n.
\end{cases}
\end{equation}
The matrix $A$ is diagonalizable with nonnegative real eigenvalues, with the minimum value being zero and nondegenerate (valid for typical nondegenerate $H_{0}$s). Additionally, it can also be seen that the off-diagonal elements of $\varrho$ evolve independently as
\begin{equation}
\varrho_{mn}(\tau)= e^{-(i\omega_{mn}+\gamma_{mn})\tau} \varrho_{mn}(0),
\label{offdiagonal}
\end{equation}
where $\omega_{mn}=E_{m}-E_{n}$ is the gap of the effective Hamiltonian and $\gamma_{mn}= (1/2) \sum_{j}\big(|l_{jm}|^2+|l_{jn}|^2\big)\geqslant 0$ represents the decay rate. Thus we obtain
\begin{align}
\varrho(\tau)=& \sum_{nm}\varrho_{nn}(0) \langle m|e^{-\tau A}|n\rangle\,|m\rangle\langle m| \nonumber\\
& + \sum_{m\neq n}e^{-(i\omega_{mn}+\gamma_{mn})\tau} \varrho_{mn}(0) |m\rangle \langle n|. 
\end{align}

\section{Fluctuation relation: Sketch of the proof}  
\label{sec:FR}

Consider a system prepared initially in a thermal state of inverse temperature $\beta_{S}$, $\varrho(0)=e^{-\beta_{S} H_{0}}/\mathrm{Tr}[e^{-\beta_{S} H_{0}}]$, which is brought into contact with a heat bath of inverse temperature $\beta_B$. Suppose that the dynamics of the system is given by the Lindblad equation (\ref{Lindblad}) and after a sufficiently long time it reaches a thermal state of the inverse temperature $\beta_{B}$, $\varrho(\infty)= e^{-\beta_{B} H_{0}}/\mathrm{Tr}[e^{-\beta_{B} H_{0}}]$. Since the initial state is diagonal in the eigenbasis of the original Hamiltonian $H_{0}$, it is evident from Eq. (\ref{offdiagonal}) that the state of the system remains diagonal in time.

The probability that the system \textit{absorbs} heat $\mathbbmss{Q}$ from the bath, in the time interval $(0,\tau)$, is given by
\begin{equation}
\label{P[+Q]}
P(+\mathbbmss{Q},\tau)=\sum_{mn}p_{m} \,p(n,\tau|m,0)\,\delta\big(\mathbbmss{Q}-[E_{n}^{(0)}-E_{m}^{(0)}]\big),
\end{equation}
where $p_{m}$ is the probability that the system is initially in the state $|m\rangle$,
\begin{equation}
\label{p[m]}
p_{m}=\mathrm{Tr}[\varrho(0)|m \rangle\langle m|]=\frac{e^{-\beta_{S} E_{m}^{(0)}}}{\mathrm{Tr}[e^{-\beta_{S} H_{0}}]}
\end{equation} 
and $p(n,\tau|m,0)$ is the probability that the system reaches the state $|n\rangle$ at time $\tau$, if it starts from the state $|m\rangle$,
\begin{align}
p(n,\tau|m,0)=\mathrm{Tr}[\varrho(\tau;m)|n \rangle\langle n|]=\langle n|e^{- \tau A}|m\rangle,
\end{align}
where $\varrho(\tau;m)$ is the state of the system at time $\tau$, if it started from the state $|m \rangle$ at time $0$ (i.e., $\varrho(0)=|m\rangle\langle m|$).

If in the transition $|m\rangle \rightarrow |n\rangle$ the system absorbs heat $\mathbbmss{Q}$ from the bath, in the reverse transition $|n\rangle \rightarrow |m\rangle$ it releases the same amount to the bath. Hence the probability that the system \textit{releases} heat $\mathbbmss{Q}$ to the bath in the time interval $(0,\tau)$ is given by
\begin{equation}
P(-\mathbbmss{Q},\tau)=\sum_{m n}p_{n}\,p(m,\tau|n,0)\,\delta\big(\mathbbmss{Q}-[E_{n}^{(0)}-E_{m}^{(0)}]\big).
\label{P[-Q]}
\end{equation}

In order to prove Eq. (\ref{Fluctuation_Relation}), we show that 
\begin{equation}
\label{ratio}
p_{m}\,p(n,\tau|m,0) =e^{\mathbbmss{Q}\Delta\beta}\,p_{n}\,p(m,\tau|n,0).
\end{equation}
First, we note that from Eq. (\ref{p[m]}) we have 
\begin{equation}
\frac{p_{m}}{p_{n}}=e^{\beta_{S} (E_{n}^{(0)}-E_{m}^{(0)})}=e^{\beta_{S}\mathbbmss{Q}}.
\label{pm/pn}
\end{equation}
The rest (and main) part of the proof is given in Appendix \ref{app:proof}, where we argue in detail that 
\begin{equation}
\frac{p(n,\tau|m,0)}{p(m,\tau|n,0)}=\frac{\langle n|e^{-\tau A}|m\rangle}{\langle m|e^{-\tau A}|n\rangle}=e^{-\beta_{B}\mathbbmss{Q}}.
\label{eq:imp}
\end{equation}
Combining this relation with Eq. (\ref{pm/pn}) completes the proof.

Having Eq. (\ref{Fluctuation_Relation}), similarly to Ref. \cite{2004-Jarzynski}, one can also obtain an upper bound on the (accumulative) probability of a heat transfer from a cold system to a hot bath ($T_{S}<T_{B}$). This can be seen from 
\begin{align}
\int_{-\infty}^{q}P(\mathbbmss{Q},\tau)\,\mathrm{d}\mathbbmss{Q} = \int_{-\infty}^{q}P(-\mathbbmss{Q},\tau)e^{\mathbbmss{Q}\Delta \beta}\,\mathrm{d}\mathbbmss{Q} \leqslant e^{q\Delta \beta}
\end{align}
applied to the case $q\leqslant 0$. This implies that the total probability of a heat transfer of amount $\geqslant|q|$ from a cold system to a hot bath drops exponentially with both $|q|$ and the temperature difference $\Delta \beta$.

\textit{Remark.---}The concepts of ``heat" and ``work" have been defined in various ways in the literature (see, e.g., Refs. \cite{Hanggi-RMP,Us-2}). However, for simplicity here we have adopted the simple definition for heat as the change of the energy of the isolated system $\mathbbmss{Q}=E_{n}^{(0)}-E_{m}^{(0)}$, a definition which sounds plausible within the weak-coupling and Markovian regime (when the system Hamiltonian does not vary in time). 

\section{Summary}
\label{sec:summary}

We have derived a fluctuation relation for the heat transfer from a system (in its thermal state) to its bath, when they are interacting such that the system would reach a unique thermal state (characterized by the temperature of the bath) through a weak-coupling, Markovian (Lindbladian) master equation. Unlike the usual fluctuation relations, where time-inverse dynamics is also assumed (microreversibility), here our relation is given by a heat transfer process and its reverse.  

\begin{acknowledgements}
M.R. thanks F. Bakhshinezhad and M. Afsary for useful discussions. A.T.R. acknowledges support by Sharif University Technology's Office of Vice President for Research and the IPM School of Nano Science.
\end{acknowledgements}

\begin{widetext}
\appendix

\section{Proof of Eq. (\ref{eq:imp})}
\label{app:proof}

We have 
\begin{equation}\label{e^(-At)}
\langle n|e^{-\tau A}|m\rangle =\sum_{s=0}^{\infty} (1/s!)(-\tau)^{s}\,\langle n| A^{s}|m\rangle.
\end{equation}
Let us expand 
\begin{align}
\label{A(p)}
\langle n|  A^{s}|m\rangle=\sum_{k_{1},k_{2},\ldots,k_{s-1}} \langle n| A|k_{1}\rangle\langle k_{1}| A|k_{2}\rangle\langle k_{2}|\ldots |k_{s-2}\rangle \langle k_{s-2}| A|k_{s-1}\rangle\langle k_{s-1}| A|m\rangle.
\end{align}
If $k_{1}\neq n \wedge k_{s-1}\neq m$, then according to Eqs. (\ref{constraint_L}) and (\ref{Matrix_A}) the right-hand side (RHS) of the above equation becomes
\begin{equation}
e^{-\beta_{B}(E^{(0)}_{n}-E^{(0)}_{m})/2}\,f_{1,1}(n,m;s),
\end{equation}
where
\begin{equation}
f_{1,1}(n,m;s)=\sum_{k_{1}\neq n,k_{2},\ldots,k_{s-2},k_{s-1}\neq m} C_{n k_{1}}C_{m k_{s-1}}e^{\beta_{B}(E^{(0)}_{k_{1}}-E^{(0)}_{k_{s-1}})/2}\,\langle k_{1}|A|k_{2}\rangle\bra{k_{2}}\ldots\ket{k_{s-2}}\langle k_{s-2}|A|k_{s-1}\rangle.
\end{equation}
Note that the function $f_{1,1}(n,m;s)$ is symmetric under $n \leftrightarrow m$, i.e., $f_{1,1}(n,m;s)=f_{1,1}(m,n;s)$.

For the case $k_{1}=n \wedge k_{s-1}\neq m$, the RHS of Eq. (\ref{A(p)}), when $k_{2}\neq n$, becomes
\begin{equation}
\label{k1,k2,kp-1}
e^{-\beta_{B}(E^{(0)}_{n}-E^{(0)}_{m})/2}\,\langle n|A|n\rangle\ \sum_{k_{2}\neq n,k_{3}\ldots,k_{s-2},k_{s-1}\neq m} C_{n k_{2}}C_{m k_{s-1}}e^{\beta_{B}(E^{(0)}_{k_{2}}-E^{(0)}_{k_{s-1}})/2}\,\langle k_{2}|A|k_{3}\rangle\bra{k_{3}}\ldots\ket{k_{s-2}}\langle k_{s-2}|A|k_{s-1}\rangle.
\end{equation}

Similarly for the case $k_{1}\neq n \wedge k_{s-1}=m$, the RHS of Eq. (\ref{A(p)}), when $k_{s-2}\neq m$, becomes
\begin{align}
e^{-\beta_{B}(E^{(0)}_{n}-E^{(0)}_{m})/2}\,\langle m|A|m\rangle\ \sum_{k_{1}\neq n,k_{2},\ldots,k_{s-1},k_{s-2}\neq m} C_{n k_{1}}C_{m k_{s-2}}e^{\beta_{B}(E^{(0)}_{k_{1}}-E^{(0)}_{k_{s-2}})/2}\,\langle k_{1}|A|k_{2}\rangle \langle k_{2}|\ldots |k_{s-3}\rangle\langle k_{s-3}|A|k_{s-2}\rangle.
\label{k1,kp-2,kp-1}
\end{align}

Adding up Eqs. (\ref{k1,k2,kp-1}) and (\ref{k1,kp-2,kp-1}) yields
\begin{equation}
e^{-\beta_{B}(E^{(0)}_{n}-E^{(0)}_{m})/2}\,f_{1,2}(n,m;s),
\end{equation}
where 
\begin{align}
f_{1,2}(n,m;s)=&\big(\langle n|A|n\rangle +\langle m|A|m\rangle \big)\sum_{k_{1}\neq n,k_{2},\ldots,k_{s-1},k_{s-2}\neq m} C_{n k_{1}}C_{m k_{s-2}}e^{\beta_{B}(E^{(0)}_{k_{1}}-E^{(0)}_{k_{s-2}})/2}\,\langle k_{1}|A|k_{2}\rangle\ \langle k_{2}|\ldots |k_{s-3} \rangle \nonumber\\
&\times  \langle k_{s-3}|A|k_{s-2}\rangle.
\end{align}
We note that, similarly to $f_{1,1}(n,m;s)$, the function $f_{1,2}(n,m;s)$ is symmetric, as $f_{1,2}(n,m;s)=f_{1,2}(m,n;s)$.

\begin{figure}[tp]
\includegraphics[scale=0.35]{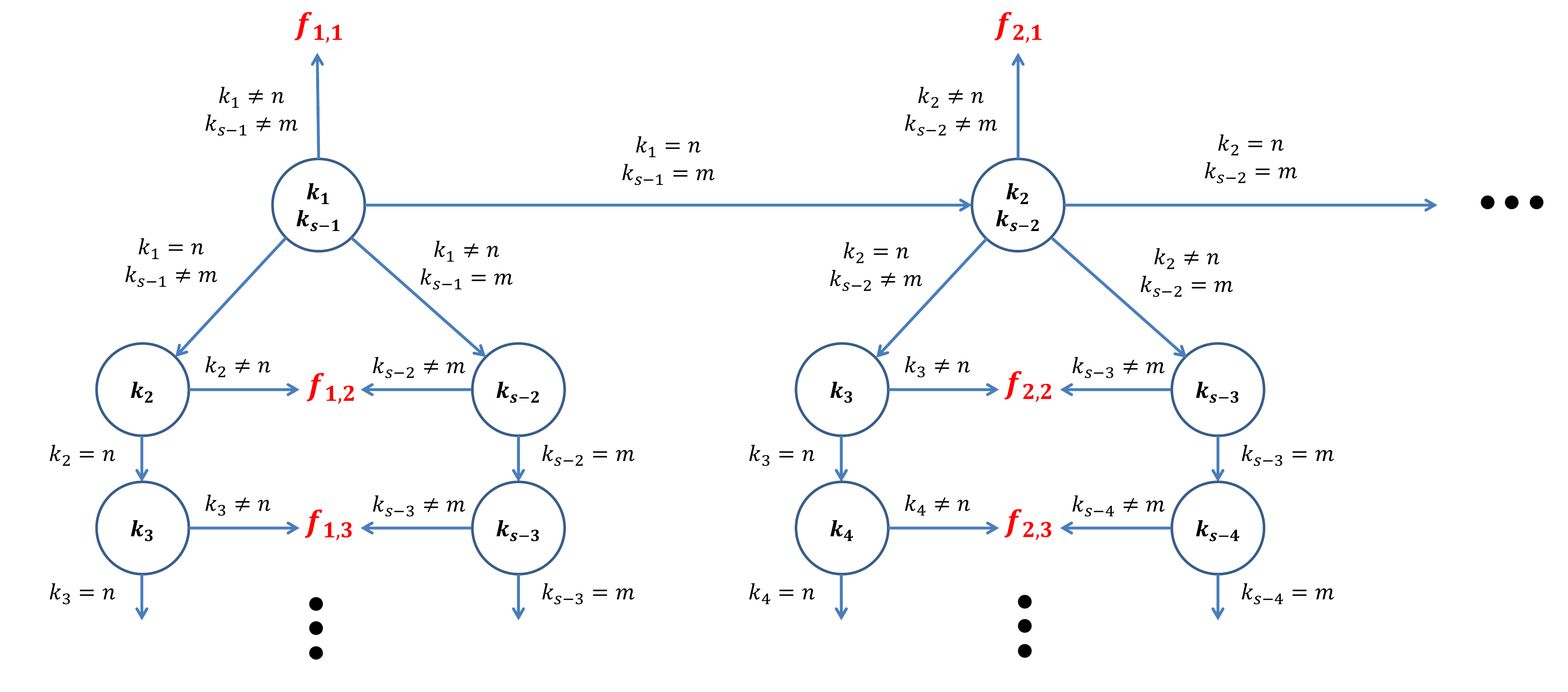} 
\caption{(Color online.) Schematic construction of the functions $f_{i,j}$.}
\label{fig:diagram}
\end{figure}

For the cases $k_{1}=n \wedge k_{2}=n \wedge k_{s-1}\neq m$ and $k_{1}\neq n \wedge k_{s-2}=m \wedge k_{s-1}=m$, we can proceed by adding the condition $k_{3}\neq n \wedge k_{s-3}\neq m$. The RHS of Eq. (\ref{A(p)}) for the sum of the cases $k_{1}=n \wedge k_{2}=n \wedge k_{3}\neq n \wedge k_{s-1}\neq m$ and $k_{1}\neq n \wedge k_{s-3}\neq m \wedge k_{s-2}=m \wedge k_{s-1}=m$ becomes
\begin{equation}
e^{-\beta_{B}(E^{(0)}_{n}-E^{(0)}_{m})/2}\,f_{1,3}(n,m;s),
\end{equation}
where
\begin{align}
f_{1,3}(n,m;s)= &\big(\langle n|A|n\rangle^{2} +\langle m|A|m\rangle^{2} \big)\sum_{ k_{1}\neq n,k_{2},\ldots,k_{s-2},k_{s-3}\neq m} C_{n k_{1}}C_{m k_{s-3}}e^{\beta_{B} (E^{(0)}_{k_{1}}-E^{(0)}_{k_{s-3}})/2}\langle k_{1}|A|k_{2}\rangle\nonumber\\
&\times \langle k_{2}|\ldots |k_{s-3}\rangle\langle k_{s-4}|A|k_{s-3}\rangle.
\end{align}

We can continue this procedure for the remaining possibilities. The same steps can be carried out for the case $k_{1}=n \wedge k_{s-1}=m$. Figure \ref{fig:diagram} summarizes the steps for calculating different terms of the summation (\ref{A(p)}). Combining all pieces, the matrix elements of $A^{s}$ is, then, given by
\begin{equation}
\label{A(p)(2)}
\langle n|A^{s}|m\rangle=e^{-\beta_{B}(E^{(0)}_{n}-E^{(0)}_{m})/2}\,\sum_{i,j} f_{i,j}(n,m;s),
\end{equation}
where
\begin{align}
f_{i,j}(n,m;s)=&(1/2)^{\delta_{j,1}}\big(\langle n|A|n\rangle\langle m|A|m\rangle\big)^{i-1}\big(\langle n|A|n\rangle^{j-1}+\langle m|A|m\rangle^{j-1}\big) \sum_{k_{1}\neq n,k_{2},\ldots,k_{s-j-2(i-1)}\neq m} C_{n k_{1}}C_{m k_{s-j-2(i-1)}}\nonumber\\
&\times e^{\beta_{B} (E^{(0)}_{k_{1}}-E^{(0)}_{k_{s-j-2(i-1)}})/2} \langle k_{1}|A|k_{2}\rangle\bra{k_{2}}\ldots \ket{k_{s-j-2i+1}}\langle k_{s-j-2i+1}|A|k_{s-j-2(i-1)}\rangle.
\end{align}

Note that each $f_{i,j}(n,m;s)$ has the $m\leftrightarrow n$ symmetry; $f_{i,j}(n,m;s)=f_{i,j}(m,n;s)$. Substituting Eq. (\ref{A(p)(2)}) in Eq.~(\ref{e^(-At)}) yields
\begin{equation}
\langle n|e^{-\tau A}|m\rangle=e^{-\beta_{B}(E^{(0)}_{n}-E^{(0)}_{m})/2}\,\sum_{s=0}\sum_{i,j=1}a_{s}(\tau)f_{i,j}(n,m;s).
\end{equation}
Since $f_{i,j}(n,m;s)$s are symmetric under the $n\leftrightarrow m$ transformation, one can conclude that
\begin{equation}
\frac{\langle n|e^{-\tau A}|m\rangle}{\langle m|e^{-\tau A}|n\rangle}=e^{-\beta_{B}(E_{n}^{(0)}-E_{m}^{(0)})}=e^{-\beta_{B}\mathbbmss{Q}}.
\end{equation}

\twocolumngrid
\end{widetext}

\end{document}